\documentclass[aps,prl,preprint,groupedaddress]{revtex4-1}

\usepackage{graphicx,amsfonts,amsmath,amssymb}

\usepackage{color}

\begin{document}


\title{Pair condensation in a dilute Bose gas with Rashba spin-orbit coupling}

\author{Rong Li, Lan Yin}
\email{yinlan@pku.edu.cn}
\affiliation{School of Physics, Peking University, Beijing 100871, China}
\date{\today}

\begin{abstract}
We show that in a two-component Bose gas with Rashba spin-orbit
coupling (SOC) two atoms can form bound states (Rashbons) with any
intra-species scattering length.  At zero
center-of-mass momentum there are two degenerate Rashbons due to
time-reversal symmetry, but the degeneracy is lifted at finite
in-plane momentum with two different effective masses.  A stable
Rashbon condensation can be created in a dilute Bose gas with
attractive intra-species and repulsive inter-species interactions.
The critical temperature of Rashbon condensation is about six times
smaller than the BEC transition temperature of an ideal Bose gas.
Due to Rashba SOC, excitations in the Rashbon condensation phase
are anisotropic in momentum space.
\end{abstract}


\maketitle {\it Introduction}. In recent several years one major
progress in ultracold atom physics was the realization of
spin-orbit coupling (SOC) in Bose-Einstein condensation \cite{Lin11} and
ultracold Fermi gases \cite{PWang,Law}.  In contrast to the
intrinsic SOC of electrons in atoms, SOC in neutral atoms refers to
the coupling between spin and center-of-mass momentum of atoms.
Bose gases with SOC have displayed very rich phase diagrams.
In experiments where the SOC is an equal-weight combination of Rashba
and Dresselhaus SOCs, several phase transitions, including magnetic
to spin-mixed states and normal to magnetics states, were observed
in Bose gases \cite{Lin11,Zhang13,Zhang12}, consistent with theoretical studies
\cite{Li}.  In a uniform Bose gas with Rashba SOC, competition between plane-wave
and spin-stripe phases were predicted \cite{CWang,Ho,Yu13}.  New phases, such as
half-vortex and skyrmion-lattice phases, were predicted in trapped systems
\cite{Wu,Hu,Sinha,Ozawa}.  In this paper, we are going to show that a stable
pairing state can appear in a two-component Bose gas with
Rashba SOC.

In contrast to the well-observed BCS-BEC crossover in Fermi gases
\cite{Regal}, pairing state of Bose gas is an exotic but never
observed phenomena.  Although Feshbach molecules of Bose atoms have been created
in experiments \cite{Donley,Xu,Thompson}, rapid particle-loss rate due to strong
inelastic collision near the resonance severely limits the molecule lifetime, making
it impossible to reach a condensed state.  Despite experimental difficulties,
the pairing state of a Bose gase has been explored theoretically \cite{Rad, Romans, Yin}, but
it was found unstable even with a weak attractive interaction away from the resonance \cite{Jeon,Basu,Yu10}.

The successful creation of SOC offers a new opportunity to realize
the pairing state in a Bose gas. As in the fermion case,
Rashba SOC changes the atom density of states (DOS) which has strong
effects on pairing and produces Rashbon \cite{Vya1}, the two-body bound
state with negative scattering length.  In the following, we first study the
two-body bound states of bosons with Rashba SOC. We find
that at zero center-of-mass momentum bound states (Rashbons) can
exist with arbitrary intra-species scattering length, while SOC
has virtually no effect on the bound state created by the
inter-species interaction. Next, we study the possibility of Rashbon
condensation in a Bose gas with Rashba SOC.  We find that Rashbon
condensation can be stabilized by a repulsive inter-species interaction.
The Rashbon condensation may be realized in a dilute Bose gas with weak
intra-species attraction and inter-species repulsion.  One signature of
this phase is the anisotropic excitation spectrum.  The Rashbon transition
temperature is about six times smaller than the ideal BEC temperature.

{\it Model}.  We consider a two-component Bose gas
with Rashba SOC, described by the Hamiltonian
$\mathcal{\hat{H}}=\mathcal{\hat{H}}_0+\mathcal{\hat{H}}_{\text{int}}$,
where
\begin{equation}
\mathcal{\hat{H}}_0=\sum_{{\bf k},\sigma}\epsilon_{{\bf k}}c^\dag_{{\bf
k}\sigma}c_{{\bf k}\sigma}+\sum_{\bf k}[S({\bf k}_{\perp})c^\dag_{{\bf k}\downarrow}c_{{\bf
k}\uparrow}+h.c.],
\end{equation}
$c_{{\bf k}\sigma}$ represents the annihilation operator of a
boson with wave-vector ${\bf k}$ and spin $\sigma$, $S({\bf
k}_{\perp})=\hbar^2 \kappa(k_x+i k_y)/m$, $\kappa$ is the strength
of Rashba SOC, ${\bf k}_{\perp}$ is the projection of ${\bf
k}$ in the x-y plane, and $\epsilon_{\bf k}=\hbar^2 k^2/2m$.  The
single-atom Hamiltonian $\mathcal{\hat{H}}_0$ can be easily diagonalized, yielding
helical excitations with energies given by $\xi_{{\bf k}\pm}=\epsilon_{\bf k}\pm\hbar^2
\kappa k_\perp/m$. The s-wave interaction between atoms is given by
\begin{equation}
\mathcal{\hat{H}}_{\text{int}}=\frac{1}{2V}\sum_{{\bf k},{\bf
k}^\prime,{\bf q};\sigma,\sigma'}g_{\sigma\sigma'}c^\dag_{{\bf
k}^\prime\sigma}c^\dag_{{\bf q}-{\bf
k}^\prime\sigma'}c_{{\bf q}-{\bf
k}\sigma'}c_{{\bf k}\sigma},
\end{equation}
where $V$ is the volume.  The inter-species coupling constants satisfy $g_{\uparrow\downarrow}=g_{\downarrow\uparrow}=4\pi \hbar^2 a'/m$, where $a'$ is the inter-species scattering length.  In the following, we consider only the symmetric case with two identical intra-species coupling constants, $g_{\uparrow\uparrow}=g_{\downarrow\downarrow}=4\pi \hbar^2 a/{m}$, where $a$ is the intra-species scattering length. In this symmetric case, the system is invariant under time-reversal transformation $({\bf k},\sigma)\rightarrow(-{\bf k},-\sigma)$.

{\it Two-body bound states}. We first study two-body bound states described by the
wave function
\begin{equation}
|\Psi\rangle_{{\bf q}}=\frac{1}{2}\sum_{{\bf k},\sigma, \sigma'}
\psi_{\sigma \sigma'}({\bf k},{\bf q}-{\bf k})c^\dag_{
{\bf k}\sigma}c^\dag_{{\bf q}-{\bf k}\sigma'}|0\rangle,
\end{equation}
where $\psi_{\sigma \sigma'}({\bf k},{\bf k}')=\psi_{\sigma'
\sigma}({\bf k}',{\bf k})$ is a coefficient. By solving the
eigenvalue problem, $H |\Psi\rangle_{\bf q} =E_{\bf q}
|\Psi\rangle_{\bf q} $, we can obtain wavefunction and eigenenergy
of bound states.  At ${\bf q}=0$, the eigenequation can be further
written as
\begin{equation}\label{eigeneq}
M_{\bf k} \psi_{\bf k}'={1 \over V}G\sum_{\bf p}\psi_{\bf p}',
\end{equation}
where $\psi_{\bf k}'$ is a four component-vector given by
$$\psi_{\bf k}'=[\psi_{\uparrow\uparrow}({\bf k},-{\bf k}),
\psi_{\downarrow\downarrow}({\bf k},-{\bf k}),\psi_{\uparrow\downarrow}({\bf k},-{\bf k}),\\
\psi_{\uparrow\downarrow}({\bf -k},{\bf k})],$$
$M_{\bf k}$ is the matrix of eigenenergy minus kinetic energy and SOC,
\begin{equation}
M_{\bf k}=\left[\begin{array}{cccc}
  \mathcal{E}_{\bf k} & 0 & S^*({\bf k}_{\perp})& -S^*({\bf k}_{\perp}) \\
   0&\mathcal{E}_{\bf k}&  -S({\bf k}_{\perp})&  S ({\bf k}_{\perp}) \\
   S({\bf k}_{\perp}) & -S^*({\bf k}_{\perp})& \mathcal{E}_{\bf k}& 0 \\
   -S({\bf k}_{\perp})& S^*({\bf k}_{\perp}) & 0 & \mathcal{E}_{\bf k}
  \end{array}\right],
\end{equation}
$\mathcal{E}_{\bf k}=E_0-2\epsilon_{\bf k}$, and G is the matrix of coupling constants,
\begin{equation}
G=\left[\begin{array}{cccc}
  g_{\uparrow\uparrow} & 0 & 0& 0 \\
  0 & g_{\uparrow\uparrow} & 0 & 0 \\
  0 & 0 & g_{\uparrow\downarrow}/2 & g_{\uparrow\downarrow}/2\\
  0 & 0 & g_{\uparrow\downarrow}/2 & g_{\uparrow\downarrow}/2
 \end{array}\right].
\end{equation}
Define the vector $Q=G \sum_{\bf k}\psi_{\bf k}'/V$, from Eq. (\ref{eigeneq}) we can
obtain an equation for $Q$,
\begin{equation}\label{eigeneqp}
Q={1 \over V}G\sum_{\bf k}M^{-1}_{\bf k} Q.
\end{equation}
Using the symmetry $S({\bf k}_{\perp})=-S(-{\bf k}_{\perp})$,
we find that
\begin{equation}
\sum_{\bf k}M^{-1}_{\bf k}=\sum_{\bf k}\det|M^{-1}_{\bf k}|\left[\begin{array}{cccc}
A_{\bf k} & 0 & 0& 0 \\
  0 & A_{\bf k} & 0 & 0 \\
  0 & 0 & A_{\bf k} & B_{\bf k}\\
  0 & 0 & B_{\bf k} & A_{\bf k}
\end{array}\right],
\end{equation}
where $A_{\bf k}=\mathcal{E}_{\bf k}^3-2\mathcal{E}_{\bf k}|S({\bf k}_{\perp})|^2$,
$B_{\bf k}=-2\mathcal{E}_{\bf k}|S({\bf k}_{\perp})|^2$, and $\det|M_{\bf k}|=\mathcal{E}_{\bf k}^2[\mathcal{E}_{\bf k}^2-4|S({\bf k}_{\perp})|^2]$.  Eq. (\ref{eigeneqp}) has three different solutions, two intra-species bound states with $Q_3=Q_4=0$ and one
inter-species bound state with $Q_1=Q_2=0$ and $Q_3=Q_4$.
Due to the symmetry $\psi_{\downarrow\uparrow}({\bf k},-{\bf k})=\psi_{\uparrow\downarrow}(-{\bf k},{\bf k})$,
the solutions always satisfy $Q_3=Q_4$ which is also guaranteed by the $G$-matrix elements $G_{34}=G_{43}=G_{33}=G_{44}$
in Eq. (\ref{eigeneqp}).
The $G$-matrix can also be chosen as a diagonal matrix with $G_{33}=G_{44}=g_{\uparrow\downarrow}$, but then the unphysical
solution with $Q_3 \neq Q_4$ has to be taken out by hand.

\begin{figure}
\centering
\includegraphics[width=0.99\columnwidth]{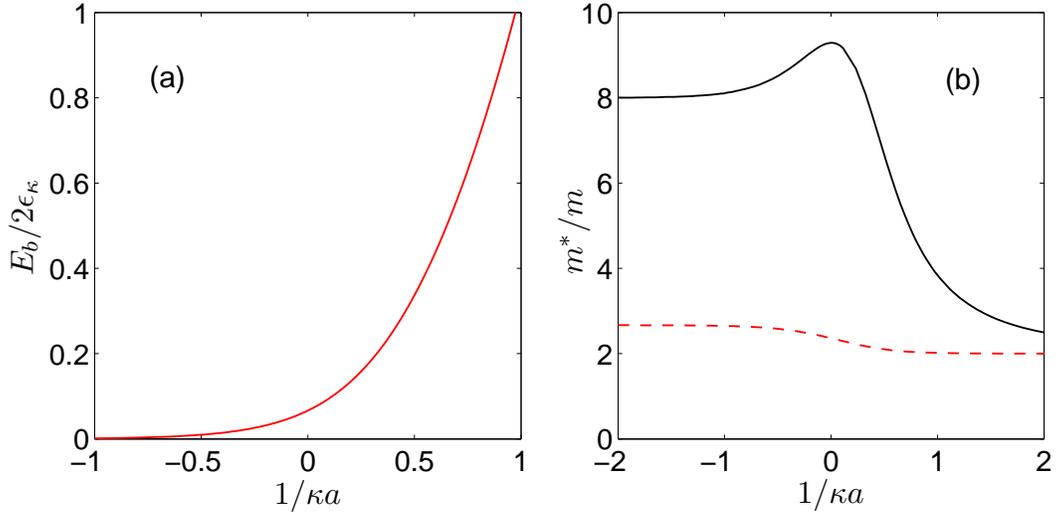}
\caption{Binding energy and effective masses of Rashbons. (a) Rashbon binding energy versus $1/(\kappa a)$ at ${\bf q}_\perp=0$.
At resonance, it is given by $E_b=0.132 \epsilon_\kappa$. (b) In the limit ${\bf q}_\perp \rightarrow 0$, two Rashbon effective masses
can be obtained from Rashbon binding energies.}
\label{fig1}
\end{figure}

For the two degenerate bound states at ${\bf q}=0$ created by the
intra-species interaction, their eigenenergy $E_0$ is determined from the equation
$1/g_{\uparrow\uparrow}=\sum_{\bf k}\det|M^{-1}_{\bf k}|A_{\bf k}/V$
which yields
\begin{equation}
\frac{m}{4\pi \hbar^2 a}=\frac{1}{2V}\sum_{\bf k}[\frac{1}{\epsilon_{\bf
k}}-\frac{1}{2\epsilon_{\bf k}-E_0}-\frac{1}{4\xi_{{\bf k}+}-2E_0}
-\frac{1}{4\xi_{{\bf k}-}-2E_0}],
\label{EEE}
\end{equation}
where the first r.h.s. term is due to $T$-matrix correction.  Eq.
(\ref{EEE}) shows that these bound states are Rashbons which can exist
with any intra-species interaction, whereas in a simple
Bose gas without SOC two-body bound states only exists in the
repulsive regime.  The binding energy defined by $E_b=-E_0-
2\epsilon_{\kappa}$ is presented in Fig.~\ref{fig1}(a) where $\epsilon_{\kappa}=\hbar^2 \kappa^2/2m$.
When the intra-species interaction is tuned from attraction to repulsion,
the binding energy monotonically increases with $1/(\kappa a)$.
We find that in the limit of $\kappa a\rightarrow 0^-$, the binding energy
has the asymptotic form
$E_b\rightarrow 8\epsilon_\kappa\exp\{4[1/(\kappa a)-1]\}$; at
resonance $1/a=0$, $E_b=0.132\epsilon_\kappa$, much smaller than
that in the fermion case \cite{Yu11}; when $\kappa a\rightarrow 0^+$,
$E_b\rightarrow\hbar^2/(m a^2)$, recovering the result of a dilute Bose
gas without SOC.

The degeneracy of Rashbons at ${\bf q}=0$ is protected
by time-reversal symmetry.  One Rashbon wavefunctions is given by
\begin{eqnarray}
 \psi_{\uparrow \uparrow }({\bf k},-{\bf k})&=& \frac{\mathcal{N}}{\mathcal{E} _{{\bf k}}}\frac{\mathcal{E} _{{\bf k}}^2-2|S({\bf k}_{\perp})|^2}{\mathcal{E} _{{\bf k}}^2-4|S({\bf k}_{\perp})|^2}, \nonumber \\
 \psi_{\downarrow \downarrow }({\bf k},-{\bf k})&=&-\frac{2 \mathcal{N}}{\mathcal{E} _{{\bf k}}}\frac{ S^2({\bf k}_{\perp})}{\mathcal{E} _{{\bf k}}^2-4|S({\bf k}_{\perp})|^2}, \nonumber \\
 \psi_{\uparrow \downarrow }({\bf k},-{\bf k})&=&-\frac{\mathcal{N}S({\bf k}_{\perp})}{\mathcal{E} _{{\bf k}}^2-4|S({\bf k}_{\perp})|^2},
 \label{intra1}
\end{eqnarray}
where $\mathcal{N}$ is
a normalization constant.
The other Rashbon wavefunction can be obtained by time-reversal
transformation
$\psi'_{\sigma \sigma'}({\bf k},-{\bf k})=\psi^*_{-\sigma -\sigma'}(-{\bf k},{\bf k})$.

Rashbon appearance in the attractive regime is due to the
increase in atom DOS at low energies by SOC.
The density of state of the lower helicity excitation $\xi_{{\bf k}-}$ is a
constant at energy minimum $\xi_{{\bf k}-}=-\epsilon_\kappa$ for $k_\perp=\kappa$ and
$k_z=0$, which leads to an infrared divergence at zero binding energy on r.s.h. of Eq.
(\ref{EEE}) and consequently Rashbon appearance
in the attractive regime.  Rashbons in the weakly attractive regime may be
helpful for experimental observation.  Since the system is far away from
resonance, the particle loss rate due to inelastic collision may be suppressed.

At ${\bf q}_\perp\neq0$, the bound-state eigenergy problem cannot be reduced to
a simple equation.   We numerically solve for bound state energies, and find that
the two Rashbons have two different effective masses, as shown in Fig.~\ref{fig1} (b).
The lift of Rashbon degeneracy is not surprising, because
two Rashbons are no longer connected by time-reversal symmetry at finite ${\bf q}_\perp$ and
the Rashbon degeneracy is no longer protected by time-reversal symmetry. The two Rashbon
effective masses behave differently with the intra-species scattering length $a$.
The bigger effective mass $m^*_+$ reaches maximum at resonance, while the
smaller effective mass $m^*_-$ decreases monotonically with
$1/(\kappa a)$.  In the limit $a \rightarrow 0^-$, we obtain
$m^*_+=8m$ and $m^*_-=8m/3$; at resonance, $m^*_+= 9.29m$ and $m^*_-=2.36m$;
in the limit $a \rightarrow 0^+$, both effective masses recover the results without SOC,
$m^*_\pm\rightarrow 2m$.

When the in-plane momentum $\hbar q_\perp$ exceeds a critical value $\hbar q_c$, the
Rashbon dissociates into excited atoms.  We find that the critical wavevector $q_c$
is different for different Rashbons, approximately satisfying the condition for Rashbon
dissociation in the effective-mass approximation, $E_0+\hbar^2q_{c\pm}^2/(2m^*_\pm)\approx-2\epsilon_\kappa$.
The critical momenta vanish in the limit of weakly
attractive interaction $a \rightarrow 0^-$, and diverge in the
opposite limit $a \rightarrow 0^+$.

For the bound state created by the inter-species interaction, its eigenenergy
at ${\bf q} = 0$ is given by
$1/g_{\uparrow\downarrow}=\sum_{\bf k}\det|M^{-1}_{\bf k}|(A_{\bf k}+B_{\bf k})/V$,
yielding
\begin{equation}
 \frac{m}{4\pi \hbar^2 a'}=\frac{1}{V}\sum_{\bf k} (\frac{1}{2\epsilon_{\bf
k}}+\frac{1}{E_0-2\epsilon_{\bf k}})
 \label{inter1}
\end{equation}
which is the same as that without SOC and gives the same result $E_b=\hbar^2/(m {a'}^2)$, whereas in the fermion case
the inter-species bound state is strongly affected by SOC \cite{Yu11}.
The wave function of this bound state is also the same as that without SOC,
\begin{eqnarray}
\psi_{\uparrow\uparrow}({\bf k},-{\bf k})&=&\psi_{\downarrow\downarrow}({\bf k},-{\bf k})=0, \nonumber \\
\psi_{\uparrow\downarrow}({\bf k},-{\bf k})&=&\frac{\mathcal{N'}}{\mathcal{E} _{\bf k}},
 \label{inter2}
\end{eqnarray}
where $\mathcal{N'}$ is a normalization factor.

The qualitative difference between Rashbon and the inter-species bound state
can be explained in terms of symmetries of their wavefunctions as given
in Eq. (\ref{intra1}) and(\ref{inter2}).  The bound state created by the
inter-species interaction consists of s-wave pairs of atoms with different
helicities, whereas in Rashbon two atoms are either with
the same helicity or p-wave symmetrized with different helicities.
In consequence, the binding energy of the bound state created by the
inter-species interaction depends on DOS of the pair energy of different helicities
$\xi_{{\bf k}+}+\xi_{{\bf k}-}=2\epsilon_{ k}$ which is independent of SOC.
Thus SOC has no effect on the binding energy of the bound
state created by the inter-species interaction.  In contrast, the Rashbon binding
energy depends on not only DOS of pair energy of different helicities, but also DOS
of pair energy of the same helicity which is half of the atom DOS with the same helicity.
The atom DOS is a constant at the lowest energy $-\epsilon_\kappa$ producing an infrared
divergence at zero binding energy on r.s.h. of Eq. (\ref{EEE}).  The change of atom DOS
by SOC is responsible for the Rashbon existence in the attractive regime.  For comparison,
in the fermionic case \cite{Vya1,Yu11}, there is no s-wave intra-species interaction due to
Fermi-Dirac statistics, and the inter-species bound state consists of p-wave pairs of atoms
with the same helicity.  The bound state is a Rashbon because of the DOS effect due to SOC.

{\it Rashbon condensation}.  Rashbons are composite bosons obeying Bose-Einstein statistics.
We consider the possibility of Bose-Einstein condensation of Rashbons
in a Bose gas with Rashba SOC.  The Rashbon
condensation can be described by pairing order parameters
$\Delta_{\uparrow\uparrow}=g_{\uparrow\uparrow}\sum_{\bf k}\langle c_{-{\bf
k}\uparrow}c_{{\bf k}\uparrow}\rangle/V$ and
$\Delta_{\downarrow\downarrow}=g_{\downarrow\downarrow}\sum_{\bf k}\langle c_{-{\bf
k}\downarrow}c_{{\bf k}\downarrow}\rangle/V$.  In general, if inter-species bound states condense,
another pairing order parameter $\Delta_{\uparrow\downarrow}=g_{\uparrow\downarrow}\sum_{\bf k}\langle c_{-{\bf
k}\uparrow}c_{{\bf k}\downarrow}\rangle/V$ needs to be introduced.  Rashbon condensation is not directly coupled
to the condensation of inter-species bound states.  In the dilute limit with weakly attractive
intra-species interaction and repulsive inter-species interaction, the Rashbon binding energy is much smaller than
the binding energy of the inter-species bound state.  In the following, we consider the system with
only Rashbon condensation and focus on the spin-balanced case, $g_{\uparrow\uparrow}=g_{\downarrow\downarrow}$ and $|\Delta_{\uparrow\uparrow}|=|\Delta_{\downarrow\downarrow}|=\Delta$.
In general there may be a phase difference between $\Delta_{\uparrow\uparrow}$ and $\Delta_{\downarrow\downarrow}$.
Without losing generality we define $\Delta_{\uparrow\uparrow}=e^{i\theta}\Delta$,
$\Delta_{\downarrow\downarrow}=e^{-i\theta}\Delta$ and $\Delta>0$.
The mean-field Hamiltonian of the Rashbon condensation phase is given by
\begin{equation}
\frac{H_{MF}}{V}=\frac{1}{2V}\sum_{\bf k}\{ B^+_{\bf k} H_{\bf k}B_{\bf
k}-2\xi_{\bf k}\}-\frac{\Delta^2}{g_{\uparrow\uparrow} }-(2g_{\uparrow\uparrow}+g_{\uparrow\downarrow})n^2,
\end{equation}
where $B^+_{\bf k}$ is the field operator with four components
$B^+_{\bf k}=[c^\dag_{{\bf k}\uparrow},c_{{\bf
-k}\uparrow},c^\dag_{{\bf k}\downarrow},c_{{\bf -k}\downarrow}]$,
$n$ is the density of each spin component,
the matrix $H_{\bf k}$ is given by
 \begin{eqnarray}\label{MFH}
H_{\bf k}&=&\left[\begin{array}{cccc}
  \xi_{\bf k} & \Delta_{\uparrow\uparrow} & S^*({\bf k}_{\perp})& 0 \\
   \Delta^*_{\uparrow\uparrow}& \xi_{\bf k} &  0 &  -S({\bf k}_{\perp}) \\
   S({\bf k}_{\perp}) & 0 & \xi_{\bf k}& \Delta_{\downarrow\downarrow} \\
   0 & -S^*({\bf k}_{\perp}) & \Delta^*_{\downarrow\downarrow} &  \xi_{\bf k}
  \end{array}\right],
\end{eqnarray}
$\xi_{\bf k}=\epsilon_{\bf k}-\mu+2g_{\uparrow\uparrow}n+g_{\uparrow\downarrow}n$, and
$\mu$ is chemical potential.

\begin{figure}
\centering
\includegraphics[width=0.95\columnwidth]{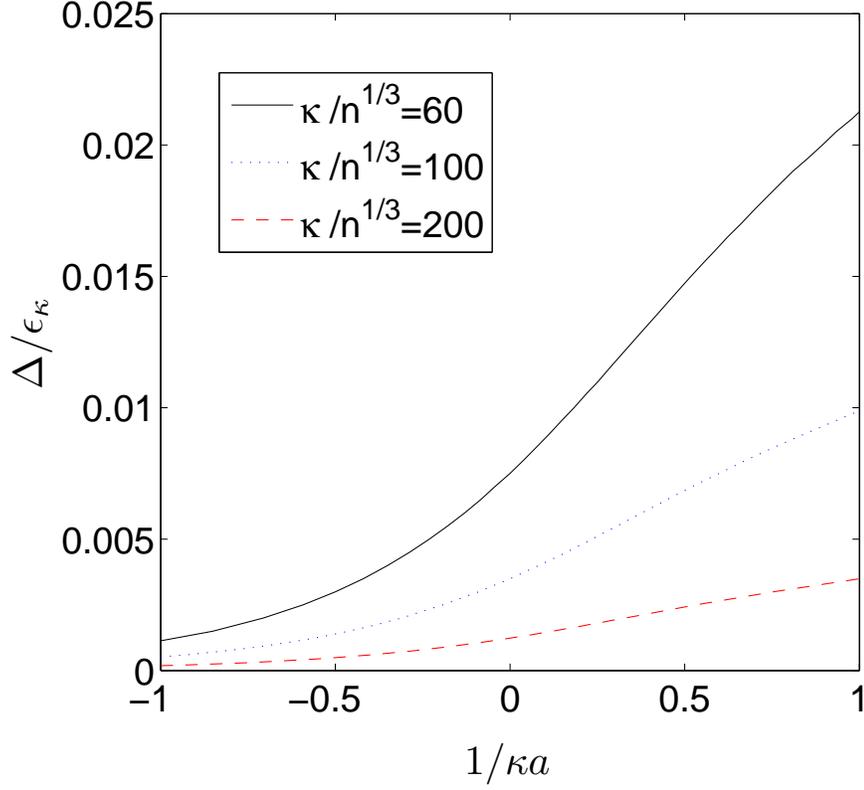}
\caption{Pairing order parameter $\Delta$ versus $1/(\kappa a)$ for different densities in the dilute limit $\kappa \gg n^{1/3}$.
For fixed $\kappa$, the order parameter $\Delta$ increases monotonously with $n$ and $1/a$. At resonance, for $ \kappa / n^{1/3}=60$,
the order parameter $\Delta=0.0075\epsilon_\kappa$ is much smaller than the binding energy $E_b=0.132\epsilon_\kappa$.}
\label{OP}
\end{figure}

The mean-field Hamiltonian Eq. (\ref{MFH}) can be diagonalized by generalized
Bogoliubov transformation.  The single-particle excitations form two branches
with excitation energies given by
\begin{equation}\label{Ek}
\varepsilon_{{\bf k}\pm}=\left[\xi^2_{\bf k}+|S({\bf k}_{\perp})|^2-\Delta^2\pm 2
|S({\bf k}_{\perp})|\sqrt{\xi^2_{\bf k}-\Delta^2\cos^2\varphi_{\bf k}}\right]^\frac{1}{2},
\end{equation}
where $\varphi_{\bf k}=\phi_{\bf k}+\theta$ and $\phi_{\bf k}=\arg(k_x+i k_y)$.
The pairing order parameters and density can be obtained
self-consistently, yielding the following equations at zero temperature
\begin{eqnarray}
\frac{1}{ g_{\uparrow\uparrow}}&=&\frac{1}{4V}\sum_{\bf k}[\frac{2}{\epsilon_{\bf k}}-\frac{1
}{\varepsilon_{{\bf k}+}}-\frac{ 1}{\varepsilon_{{\bf k}-}}-\frac{|S({\bf k}_{\perp})|\cos^2\varphi_{\bf k}
}{\sqrt{\xi^2_{\bf k}-\Delta^2\cos^2\varphi_{\bf
k}}}(\frac{1 }{\varepsilon_{{\bf k}+}}-\frac{1}{ \varepsilon_{{\bf k}-}})], \label{Gap}\\
n&=&\frac{1}{4V}\sum _{\bf k} [\frac{\xi _{\bf k} }{\varepsilon_{{\bf k}-}}(1-\frac{|S({\bf k}_{\perp})|}{\sqrt{\xi^2_{\bf k} -\Delta^2\cos^2\varphi_{\bf k} }})\nonumber +\frac{\xi _{\bf k} }{\varepsilon_{{\bf k}+}}(1+\frac{|S({\bf k}_{\perp})|}{\sqrt{\xi^2_{\bf k} -\Delta^2\cos^2\varphi_{\bf k} }})-2].
\label{Num}
\end{eqnarray}
We numerically solve Eq. (\ref{Gap}) and find that the mean-field solution always exist in the dilute limit $n\rightarrow 0$,
as shown in Fig. \ref{OP}.

\begin{figure}
\centering
\includegraphics[width=0.99\columnwidth]{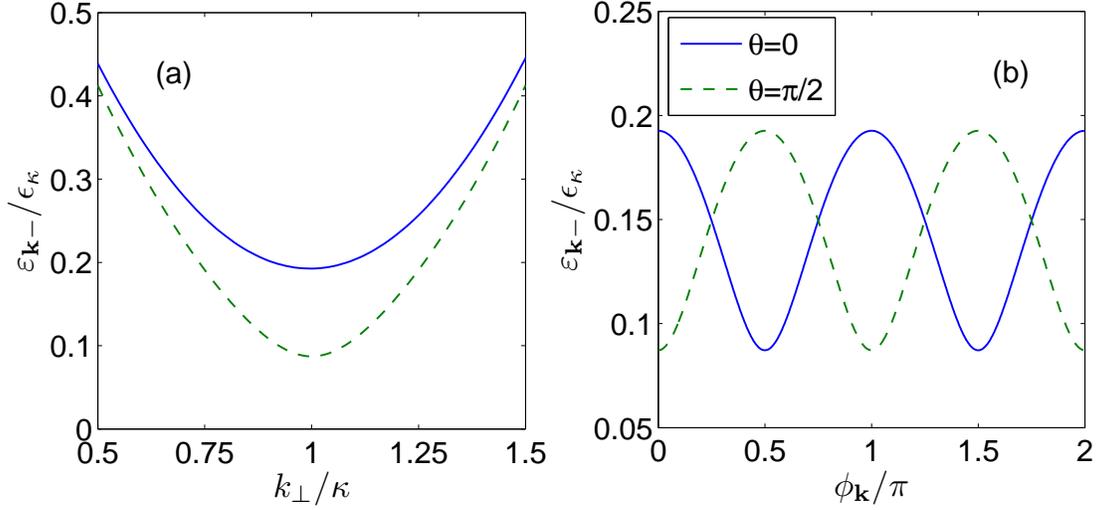}
\caption{Anisotropy of lower quasi-particle excitation energy
$\varepsilon_{{\bf k}-}$ at $\mu'/\epsilon_\kappa=-1.2$,
$\Delta /\epsilon_\kappa=0.18$ and $k_z=0$, where $\mu'=\mu-2g_{\uparrow\uparrow}n-g_{\uparrow\downarrow}n$. (a) $\varepsilon_{{\bf k}-}$ along
x-axis (solid line) and y-axis (dash line) are plotted as functions
of $k_\perp/\kappa$ for $\theta=0$.
The anisotropy is stronger at low
energies and weaker at higher energies. (b) $\varepsilon_{{\bf k}-}$ at
$k_\perp=\kappa$ versus $\phi_{\bf k}$ for different $\theta$.} \label{Three}
\end{figure}

In Rashbon condensation, quasi-particle excitation energies given in Eq. (\ref{Ek})
are anisotropic, dependent on the angle $\varphi_{\bf k}=\phi_{\bf k}+\theta$.
This anisotropy is stronger at low energies when
$k_\perp$ is near $\kappa$, as shown in Fig. \ref{Three}(a).  At higher
energies, the anisotropy becomes weaker and eventually disappears.
This anisotropic effect is caused by the coupling between
pairing order parameters $\Delta _{\uparrow\uparrow}$ and $\Delta_{\downarrow\downarrow}$
due to SOC.  For a spin-up atom with wave-vector ${\bf k}$,
SOC can flip its spin down with a phase $\phi_{\bf k}$.
This phase becomes $\phi_{\bf k}+\pi$ for the spin-up atom with
opposite wave-vector $-{\bf k}$.  These two spin-flips can turn an atom pair from total spin-up
to total spin-down states with phase $2\phi_{\bf k}+\pi$.  If
$2\phi_{\bf k}+\pi+2\theta=2l\pi$ where $l$ is an integer, spin-flips are encouraged and
the quasi-particle energy $\varepsilon_{{\bf k}-}$ is
at minimum.  If $2\phi_{\bf k}+2\theta=2l\pi$, spin-flips are discouraged
and the quasi-particle energy is at maximum.  As shown in Fig.
\ref{Three}(b), the quasi-particle energy $\varepsilon_{{\bf k}-}$ shows a periodic behavior as a function
of $\phi_{\bf k}$ with period $\pi$.

In the following, we focus on Rashbon condensation in the dilute
limit with attractive intra-species interaction, $\kappa \gg n^{1/3}$  and $(-a)^{-1} \gg n^{1/3}$.
Since in the dilute limit the distance between Rashbons is the largest length scale, the structure of Rashbons is not affected by the weak interaction between Rashbons, which is very similar to the BEC limit of BEC-BCS crossover in Fermi gases.
In this limit, Eq. (\ref{Gap}) can be solved analytically, and we find that the order
parameter $\Delta $ is much smaller than Rashbon binding energy,
$$\Delta \approx 4\sqrt{2\pi}(\frac{n}{\kappa ^{3}})^\frac{1}{2}(\epsilon_\kappa E_b)^\frac{1}{2}\ll E_b.$$
The attractive intra-species interaction tends to make the system unstable.  If the Rashbon
condensation is stable, the positive compressibility condition $\partial \mu/\partial n>0$
must be satisfied.  We find that in the dilute limit this stability condition
is given by $\kappa(a'+2a)>3/2$.  Therefore a repulsive inter-species
interaction with $\kappa>3/(2a'+4a)\gg n^{1/3}$ is required to
stabilize the Rashbon condensation in a dilute Bose gas with Rashba SOC.

In the Rashbon condensation phase, in addition to single-particle excitations,
there are also pair excitations.  At the transition temperature $T_c$ of Rashbon condensation,
pair excitations are quadratically dispersed.  In the dilute limit with attractive
intra-species interaction, they have effective masses approximately as same as those of Rashbons in vacuum.
Since in this limit the Rashbon binding energy is much bigger than $k_B T_c$,
single-particle excitations can be neglected at $T_c$, and only excited Rashbons contribute
to the density at $T_c$,
\begin{equation}\label{Den}
n=\frac{1}{V}\sum_{{\bf q},s}{'}\frac{1}{e^{\beta (E_{{\bf
q}s}-E_0)}-1},
\end{equation}
where $s=\pm$, $E_{{\bf q}\pm} \approx E_0+\hbar^2q^2_z/(4m)+\hbar^2q^2_\perp/(2m^*_\pm)$ are Rashbon energies in the
effective mass approximation, and $\sum{'}$ denotes the summation over ${\bf q}$ for $|q_\perp| \leq q_{c\pm}$.
From Eq. (\ref{Den}), we obtain the transition temperature
\begin{equation}\label{Tc}
T_c=[\sqrt{2}(m^*_++m^*_-)/m]^{-\frac{2}{3}}T_a\approx 0.164 T_a,
\end{equation}
where $T_a=2\pi \zeta^{-\frac{2}{3}}(\frac{3}{2})\hbar^2n^{2/3}/(k_B m)$ is the critical temperature of an ideal Bose gas
and $\zeta(x)$ is the Riemann zeta function.  Eq. (\ref{Tc}) shows that
the transition temperature of Rashbon condensation $T_c$ in the dilute limit
is about six times smaller than
the BEC transition temperature of an ideal Bose gas.

In current experiments in $^{87}$Rb, the strength of Rashba SOC is limited by the wavelength of the Raman laser $\lambda=804.1$nm, $\kappa\leq7.8\times10^6$ m$^{-1}$ \cite{Lin11}.  With background intra-species scattering length $a_{bg}=100a_0$ and density of the order of $10^{13}$ cm$^{-3}$ \cite{Long13}, the dilute region of Rashbon condensation is hardly reachable.  With the new proposal to generate Rashba SOC \cite{And13, Ken13}, if $\kappa$ can be enhanced to $2\times10^8$ m$^{-1}$ and scattering lengths can be tuned to $a=-95a_0$ and $a'>330a_0$, Rashbon condensation may be observed around 29nK with $n=10^{13}$ cm$^{-3}$ in $^{87}$Rb.

{\it Discussion and conclusion.}
We have shown that Rashbon condensation can be mechanically stable in a dilute Bose gas with Rashba SOC and weakly attractive intra-species interaction.
In this dilute region, we expect that the particle loss rate is suppressed because of its density dependence.  As shown in experiments on $^{85}$Rb in the dilute region \cite{Donley, Thompson},
the loss rate of Feshbach molecules is much smaller than the molecule binding energy.  Now with the help of Rashba SOC, the Rashbon binding energy is exponentially small,
and the lifetime of dilute Rashbon condensation is expected to be long enough for experimental observations.

There are a lot of interesting questions to be answered
about Rashbon condensation.  Collective excitations in this  phase
are worth to explore. Another important question is whether or not at a higher density there is
a quantum phase transition between Rashbon condensation and mixture of atom and
Rashbon condensates.  We plan to address these issues in future studies.

In summary, we find that two Bose atoms with Rashba SOC can form a
Rashbon with any intra-species interaction.  In contrast, the bound state
created by the inter-species interaction is not affected by SOC.
At zero center-of-mass momentum there are two degenerate Rashbons with the degeneracy protected by
time-reversal symmetry.  The degeneracy is lifted at finite in-plane momentum
with two different effective masses.  We explore the possibility of Rashbon
condensation in a dilute Bose gas with Rashba SOC and attractive intra-species interaction.
We find that Rashbon condensation can be stabilized by a repulsive inter-species interaction.
In Rashbon condensation, the single-particle excitation energy is anisotropic,
due to coupling between pairing order parameters by SOC.  The transition temperature
of Rashbon condensation is about six times smaller than
that of BEC in an ideal Bose gas.

{\bf Acknowledgement}. We would like to thank Z. Q. Yu, W. Zhang, and T.-L. Ho for helpful discussions.  This work is supported by NSFC under Grant No 11274022.

\end{document}